\documentstyle[12pt]{article}
\newlength{\pubnumber} 

\catcode`\@=11
\@addtoreset{equation}{section}
\def\section{\@startsection{section}{1}{\z@}{3.5ex plus 1ex minus .2ex}
 {2.3ex plus .2ex}{\large\bf}}
\def\subsection{\@startsection{subsection}{2}{\z@}{2.3ex plus .2ex}
 {2.3ex plus .2ex}{\bf}}

    \renewcommand{\baselinestretch}{1.4}
\begin{document}

\begin{titlepage}
\samepage{
\setcounter{page}{1}
\rightline{UFIFT-HEP-96-31}
\rightline{\tt hep-ph/9612400}
\rightline{December 1996}
\vfill
\begin{center}
 {\Large \bf  R-parity violation in\\
            superstring derived models\\}
\vfill
 {\large Alon E. Faraggi\footnote{
   E-mail address: faraggi@phys.ufl.edu}\\}
\vspace{.12in}
 {\it Institute For Fundamental Theory, \\
  University of Florida, \\
  Gainesville, FL 32611 USA\\}
\end{center}
\vfill
\begin{abstract}
  {\rm
The ALEPH collaboration has recently reported a significant excess of 
four--jet events at the LEP 1.5 and LEP 2 experiments. While not yet 
confirmed by the other collaborations, it was recently proposed that
this excess may be explained by certain R-parity violating operators 
in supersymmetric models. R--parity violating operators introduce the
danger of inducing rapid proton decay. I discuss how the operators required
to explain the ALEPH four jet events may arise from superstring derived
models without inducing rapid proton decay. 
}
\end{abstract}
\vfill
\smallskip}
\end{titlepage}

\setcounter{footnote}{0}

\def\beq{\begin{equation}}
\def\eeq{\end{equation}}
\def\beqn{\begin{eqnarray}}
\def\eeqn{\end{eqnarray}}
\def\AEF{A.E. Faraggi}
\def\NPB#1#2#3{{\it Nucl.\ Phys.}\/ {\bf B#1} (19#2) #3}
\def\PLB#1#2#3{{\it Phys.\ Lett.}\/ {\bf B#1} (19#2) #3}
\def\PRD#1#2#3{{\it Phys.\ Rev.}\/ {\bf D#1} (19#2) #3}
\def\PRL#1#2#3{{\it Phys.\ Rev.\ Lett.}\/ {\bf #1} (19#2) #3}
\def\PRT#1#2#3{{\it Phys.\ Rep.}\/ {\bf#1} (19#2) #3}
\def\MODA#1#2#3{{\it Mod.\ Phys.\ Lett.}\/ {\bf A#1} (19#2) #3}
\def\IJMP#1#2#3{{\it Int.\ J.\ Mod.\ Phys.}\/ {\bf A#1} (19#2) #3}
\def\nuvc#1#2#3{{\it Nuovo Cimento}\/ {\bf #1A} (#2) #3}
\def\etal{{\it et al,\/}\ }
\hyphenation{su-per-sym-met-ric non-su-per-sym-met-ric}
\hyphenation{space-time-super-sym-met-ric}
\hyphenation{mod-u-lar mod-u-lar--in-var-i-ant}


\setcounter{footnote}{0}

Particle physics has for long awaited the experimental deviation from the 
Standard Model predictions that will guide the way to the physics
beyond the Standard Model. Such a discovery may be just around the 
corner. Recently, the ALEPH collaboration has reported an excess 
in the four--jet event cross section which is several sigmas
above the Standard Model prediction \cite{alephone,alephtwo}.
Perhaps even more intriguing is the 
fact that both at LEP 1.5 and LEP 2 runs ALEPH has observed a sharp 
peak at $106.1\pm0.8$ GeV, corresponding to 18 events with 3.1 expected 
from QCD background. The di--jet mass difference distribution 
of the selected 18 events is consistent with a value around 10 GeV. 
If interpreted as a particle pair production this together with
the information on the di--jet mass sum suggests that the
two particle produced have masses of about 58 and 48 GeV and production
of same mass particles is disfavored. By extracting information 
on the primary parton \cite{alephthree},
it is concluded that the pair produced 
particles have a sizable charge and that neutral particle production
is disfavored \cite{alephone,alephtwo}. 
Absence of $b$--quarks in the final states disfavors
the hypothesis of Higgs--boson production. 

Recently, it was proposed \cite{giudice} that the ALEPH excess of four 
jet events can be explained in supersymmetric models with
R--parity violation \cite{rparityv}. 
According to this proposal left--handed
and right--handed selectrons ${\tilde e}_L{\tilde e}_R$
are pair produced at LEP. The selectron pair then decay further
by the R--parity violating operator
\begin{equation}
\lambda_{ijk} L^iQ^j{d}^k
\label{rxloperator}
\end{equation}
where the standard notation for lepton and quark superfields has
been used and $i,j,k$ denote the generation indices, and 
absence of top or bottom quarks in the final states restricts 
only the $\lambda_{1jk}$ with $j,k=1,2$ to be nonzero \cite{giudice}. 

It is well known however that R--parity violation may induce
rapid proton decay. If in addition to the operator in Eq. (\ref{rxloperator})
one also has the operator 
\begin{equation}
\eta_{ijk}u^id^jd^k
\label{rxboperator}
\end{equation}
with unsuppressed $\eta_{ijk}$ couplings then the proton decays rapidly.
A quick estimate of the constraint from proton lifetime gives
\begin{equation}
(\lambda\eta)<g^2\left({M_{\rm squark}\over{M_{\rm GUT}}}\right)^2
\label{constraint}
\end{equation}
Therefore, if $\lambda_{1jk}\sim10^{-4}$ as proposed in ref. 
\cite{giudice}, and taking $M_{\rm squark}\sim1$ TeV 
$M_{\rm GUT}\sim10^{16}$ GeV, naively requires $\eta<10^{-22}$.
Thus, proton constraints essentially requires $\eta\equiv0$
if the R--parity violation is to explain the excess of
ALEPH four jet events. 

Therefore, the problem is to understand why the couplings
in Eq. (\ref{rxloperator}) are allowed while the couplings
in Eq. (\ref{rxboperator}) are forbidden. 
In this paper I discuss this problem in the context of
realistic superstring derived models. 

To study this problem I examine the superstring models
which are constructed in the free fermionic formulation
\cite{fff,revamp,fny,alr,slm,eu,gcu,custodial}. 
This class of superstring models reproduce many of the 
properties of the Standard Model, like three chiral generations
with the Standard Model gauge group and existence of Higgs doublets
which can generate realistic fermion mass spectrum. Two 
of the important features in this class of models
is the existence of a stringy doublet--triplet splitting mechanism 
which resolves the GUT hierarchy problem \cite{ps}
and the fact that the chiral generations all fall
into the 16 of $SO(10)$. This last 
property admits the standard embedding of the weak
hypercharge in $SO(10)$ and is crucial for the agreement of these 
models with $\sin^2\theta_W(M_Z)$ and 
$\alpha_{\rm strong}(M_Z)$ \cite{gcu,df}. 

The superstring models under consideration are
constructed in two steps. In the first step the
observable gauge symmetry is broken to $SO(10)\times SO(6)^3$.
There are 48 generations in the chiral 16 representation
of $SO(10)$ with $N=1$ space--time supersymmetry.
In the second step the $SO(10)$ symmetry is broken to one of its
subgroups, $SU(5)\times U(1)$, $SO(6)\times SO(4)$ or 
$SU(3)\times SU(2)\times U(1)^2$. The flavor $SO(6)^3$
symmetries are broken to $U(1)^n$, where $n$ may vary
between 3--9, and the number of generations is reduced to three. 
The symmetry is then broken further in the effective field 
theory and the weak hypercharge is some linear combination of 
the Cartan subgenerators. For example, in the standard--like models
the weak hypercharge is given by, 
\begin{equation}
U(1)_Y={1\over3}U(1)_C+{1\over2}U(1)_L
\label{forexample}
\end{equation}
The chiral generations in these superstring models
are obtained from the 16 multiplets of $SO(10)$ and carry 
charges under the flavor symmetries. These models typically 
contain an ``anomalous'' $U(1)$ symmetry which requires
that some fields in the massless string spectrum obtain 
non--vanishing VEVs \cite{dsw}. Further details on the
construction of the realistic free fermionic models are given 
in ref. \cite{slm}.

In general in string models one expects the appearance 
of R--parity violating terms of the form of Eq. (\ref{rxloperator})
and (\ref{rxboperator}). If both are not suppressed then the proton
decays much too fast. If the $B-L$ generator is gauged like 
in $SO(10)$ then these terms are forbidden at the cubic level by 
gauge invariance. However, they may still be generated
from nonrenormalizable terms that contain the right--handed 
neutrino. 
\begin{equation}
\eta_1(uddN)\Phi+\eta_2(QLdN)\Phi.
\label{quartic}
\end{equation}
where $\Phi$ is a combination of fields that
fixes the string selection rules and gets a VEV
of $O(M_{Pl})$ and $N$ is the Standard Model singlet
in the 16 of $SO(10)$. Thus, the ratio $\langle N\rangle/M_{Pl}$
controls the rate of proton decay. In general, 
terms of the form of Eq. (\ref{quartic}) are expected 
to appear in string models at different orders of nonrenormalizable
terms. For example, in the model of ref. \cite{eu} such terms 
appear at order $N=6$
\begin{eqnarray}
 &(u_3d_3+Q_3L_3)d_2N_2\Phi_{45}{\bar\Phi}_2^{-}\nonumber\\
+&(u_3d_3+Q_3L_3)d_1N_1\Phi_{45}\Phi_1^{+}\nonumber\\
+&u_3d_2d_2N_3\Phi_{45}{\bar\Phi}_2^{-}+
  u_3d_1d_1N_3\Phi_{45}\Phi_1^{+}\nonumber\\
+&Q_3L_1d_3N_1\Phi_{45}\Phi_3^+
+Q_3L_1d_1N_3\Phi_{45}\Phi_3^+\nonumber\\
+&Q_3L_2d_3N_2\Phi_{45}{\bar\Phi}_3^-
+Q_3L_2d_2N_3\Phi_{45}{\bar\Phi}_3^-.
\label{ordersix}
\end{eqnarray}
In this model the states from the sector $b_3$ are identified 
with the lightest generation.
It is therefore seen that if any of $N_1$, $N_2$ or $N_3$
gets a Planck scale VEV, dimension four operators
may be induced which would result in rapid proton decay.
It is interesting to note that all the terms in Eq. (\ref{ordersix})
contain the field $\Phi_{45}$. If the VEV of $\Phi_{45}$ vanishes
then all the higher order terms are identically zero.
In this specific model due to the anomalous $U(1)$ symmetry
$\Phi_{45}$ must get a VEV and, in general, dimension four
operators may be induced. Nevertheless, this observation
suggests the possibility that slight variation of the model
will result in a field appearing in these terms which is not required 
to get a VEV. However, even if such a possibility can work
we see that both the desired terms of the form $QLd$ and
the undesired terms of the form $udd$ are induced, or forbidden, simultaneously. 

In the flipped $SU(5)$ model similar terms may arise
from the terms 
\begin{equation}
FF{\bar f}H\Phi^n.
\label{sufive}
\end{equation} 
Here $F$ and $H$ are in the $(10,1/2)$ representation 
and ${\bar f}$ is in the $({\bar 5},-3/2)$ representation
of $SU(5)\times U(1)$. The field $F$ contains the $Q$, $d$, $N$
fields and ${\bar f}$ contains the $u$ and $L$ fields. The Standard
Model singlet, $N$ in the Higgs field $H$ obtains a VEV which 
breaks the $SU(5)\times U(1)$ symmetry to the Standard Model
symmetry. Thus, terms of the form of Eq. (\ref{sufive})
produce simultaneously the terms in Eq. (\ref{rxloperator}) and
Eq. (\ref{rxboperator}). Terms of the form of Eq. (\ref{sufive})
are in general found in the string models \cite{elnone}.
Therefore, to produce only the terms
of the form of Eq. (\ref{rxloperator}) while preventing the 
terms in Eq. (\ref{rxboperator}) requires a different mechanism.

In the case of the $SO(6)\times SO(4)$ superstring models
the Standard Model fermions are embedded in the
\begin{eqnarray}
&F_L\equiv(4,2,1)=Q+L\nonumber\\
&{\bar F}_R\equiv({\bar 4},1,2)=u+d+e+N
\label{psleft}
\end{eqnarray}
representations of the $SU(4)\times SU(2)_L\times SU(2)_R$.
Note that $F_L+{\bar F}_R$ make up the 16 spinorial representation
of $SO(10)$. The dangerous dimension four operators are obtained 
in this case from the operator 
\begin{equation}
F_LF_L{\bar F}_R{\bar H}_R~~{\rm and}~~
{\bar F}_R{\bar F}_R{\bar F}_R{\bar H}_R
\label{d4so64}
\end{equation}
where ${\bar H}_R$ is the Higgs representation 
which breaks the extended non--Abelian symmetry.
We observe that in the $SO(6)\times SO(4)$ type models,
like the $SU(3)\times SU(2)\times U(1)^2$ type models,
the operator in Eqs. (\ref{rxloperator}) and (\ref{rxboperator})
arise from two distinct operators. 

Next, I turn to the model of Ref. \cite{custodial}. 
The detailed spectrum of this model and the quantum numbers 
are given in Ref. \cite{custodial}. 
In this model the observable gauge group formed by the gauge bosons
from the Neveu--Schwarz sector alone is
\begin{equation}
SU(3)_C\times SU(2)_L\times U(1)_C\times U(1)_L\times U(1)_{1,2,3,4,5,6}
\label{nsgaugebosons}
\end{equation}
However, in this model two additional gauge bosons appear 
from the twisted sector ${\bf 1}+\alpha+2\gamma$.
These new gauge bosons are singlets of the non--Abelian gauge 
group but carry $U(1)$ charges. Referring to this generators 
as $T^{\pm}$, then together with the linear combination
\begin{equation}
T^3\equiv{1\over4}\left[U(1)_C+U(1)_4+U(1)_5+U(1)_6+U(1)_7-U(1)_9\right]
\label{t3}
\end{equation}
the three generators $\{T^3,T^{\pm}\}$ together form an enhanced 
$SU(2)_{\rm custodial}$ symmetry group. 
Thus, the original observable symmetry group is enhanced to 
\begin{equation}
       SU(3)_C \times SU(2)_L \times SU(2)_{\rm cust} \times
             U(1)_{C'} \times U(1)_L
           \times U(1)_{1,2,3} \times U(1)_{4',5',7''}
\end{equation}
The different combinations of the $U(1)$ generators are given in ref. \cite{custodial,df}. The weak hypercharge is still defined 
as a combination of $U(1)_C$ and $U(1)_L$. However in the present 
model $U(1)_C$ is part of the extended $SU(2)_{\rm custodial}$
symmetry. We can express $U(1)_C$ in terms of the 
new orthogonal $U(1)$ combinations, 
\begin{equation}
    {1\over3}\,U(1)_{C}~=~{2\over5}\biggl\lbrace U(1)_{C^\prime}+
     {5\over{16}}\,\biggl\lbrack T^3+{3\over5}\,U_{7^{\prime\prime}}
\biggr\rbrack
        \biggr\rbrace~.
\label{U1Cin274}
\end{equation}
and the weak hypercharge is given as before by
the linear combination
\begin{equation}
U(1)_Y={1\over3}U(1)_C+{1\over2}U(1)_L
\label{weakhypercharge}
\end{equation}
The weak hypercharge depends on the diagonal generator
of the custodial $SU(2)$ gauge group. 
We can therefore instead define the new linear combination with this term
removed,
\begin{eqnarray}
        U(1)_{Y'} &\equiv& U(1)_Y - {1\over 8}\,T^3 \nonumber\\
        &=& {1\over 2}\,U(1)_L + {5\over{24}}\,U(1)_C \nonumber\\
         &&~~~~~-{1\over8}\,\biggl\lbrack
U(1)_4+U(1)_5+U(1)_6+U(1)_7-U(1)_9\biggr\rbrack~,
\label{U1pin274}
\end{eqnarray}
so that the weak hypercharge is expressed in terms of $U(1)_{Y'}$ as
\begin{equation}
        U(1)_{Y} = U(1)_{Y'} + {1\over 2}\,T^3 ~~~~\Longrightarrow~~~~
           Q_{\rm e.m.} = T^3_L + Y = T^3_L + Y' + {1\over 2}\,T^3_{\rm
cust}~.\label{Qemin274}
\end{equation}
The final observable gauge group then takes the form
\begin{equation}
       SU(3)_C \times SU(2)_L \times SU(2)_{\rm cust}\times U(1)_{Y'} ~\times
        ~\biggl\lbrace ~{\rm seven~other~}U(1){\rm ~factors}~\biggr\rbrace~.
\label{finalgroup}
\end{equation}
These remaining seven $U(1)$ factors must be chosen
as linear combinations of the previous $U(1)$ factors so as to be orthogonal
to the each of the other factors in (\ref{finalgroup}).

The full massless spectrum of this model is given in Ref. \cite{custodial}.
In this model the charged and neutral leptons transform as doublets of the 
$SU(2)_{\rm custodial}$ symmetry while the quarks are singlets.
Therefore, because of the custodial $SU(2)$ symmetry
the terms of the form
\begin{equation}
QLdN
\label{qldn}
\end{equation}
are invariant under the custodial $SU(2)$ symmetry, while the 
terms of the form 
\begin{equation}
uddN
\label{uddn}
\end{equation}
are not invariant. We could contemplate tagging another $N$ field
to Eq. (\ref{uddn}) which will render it invariant under
$SU(2)_{\rm custodial}$. However, this will spoil the invariance 
under $U(1)_L$. We therefore find that the baryon number violating 
operators, Eq. (\ref{uddn}) vanish to all orders in the model of
ref. \cite{custodial}. Therefore, this model admits 
the type of custodial symmetries which allow the 
R--parity lepton--number violating operators of Eq. (\ref{rxloperator})
while they forbid the baryon--number violating operators of
Eq. (\ref{rxboperator}). This conclusion was verified
by an explicit search of nonrenormalizable terms up to order
$N=10$. On the other hand we find already at order $N=6$ the 
non--vanishing terms
\begin{eqnarray}
&Q_1d_3L_3N_1\Phi_{45}\Phi_1~+~Q_1d_2L_2N_2\Phi_{45}{\bar\Phi}_{45}\nonumber\\
&Q_1d_1L_3N_3\Phi_{45}\Phi_1~+~Q_1d_1L_2N_2\Phi_{45}{\bar\Phi}_{45}\nonumber\\
&Q_2d_3L_3N_2{\Phi_{45}}{\bar\Phi}_2+
 Q_2d_2L_1N_1\Phi_{45}{\bar\Phi}_{45}\nonumber\\
&Q_2d_2L_3N_3\Phi_{45}{\bar\Phi}_2~+~
 Q_2d_1L_1N_2\Phi_{45}{\bar\Phi}_{45}
\label{model7}
\end{eqnarray}
At higher orders additional terms will appear. 
It is therefore seen that while the R--parity baryon number violating 
operators are forbidden to all orders of nonrenormalizable terms the 
lepton number violating operators are allowed. This is precisely what
is required if the R--parity violation interpretation of the excess of
four jet events observed by the ALEPH collaboration is correct. 

Let us note some further remarks with in regard to the model
proposed in Ref. \cite{giudice}. As claimed there the R--parity 
interpretation prefers low values of $\tan\beta$ and therefore to
allow perturbative unification requires some intermediate thresholds. This
is precisely the scenario suggested by the class of superstring 
standard--like models \cite{top}. In this class of models the top--bottom
quarks mass hierarchy arises due to the fact that only the 
top quark gets its mass from a cubic level term in the superpotential
while the bottom quark gets its mass term from a higher order term. 
Thus, in this class of models the top--bottom mass splitting arises
due to a hierarchy of the Yukawa couplings rather than a large value of
$\tan\beta$. It has similarly been proposed in the context
of this class of superstring models that intermediate matter
thresholds are required for resolution of the string scale gauge
coupling unification problem \cite{gcu,df}. 

To conclude, it was shown in this paper that string models
can give rise to dimension four R--parity lepton number violating operators
while forbidding the baryon number violating operators. 
Thus, R--parity violation is allowed while proton decay
is forbidden. It will be of further interest to examine whether
similar mechanism can operate in other string 
models \cite{lykken,ibanez}. For example,
the $SO(6)\times SO(4)$ type models are of particular interest
as they also can in principle differentiate between the lepton--number and 
baryon--number violating operators. It is of further interest to
study whether the string models can actually give sizable R--parity
violation which is not in conflict with any observation.
Finally, we eagerly await 
the experimental resolution of the observed excess in the ALEPH four
jet events. 


It is a pleasure to thank G. Giudice for valuable discussions
and the CERN theory group for hospitality.
This work was supported in part by DOE Grant 
No.\ DE-FG-0586ER40272.

\bibliographystyle{unsrt}

\begin{thebibliography}{99}
\bibitem{alephone} D. Buskulic {\it et al.}, (ALEPH Coll.), 
			{\it Z. Phys.}\/ {\bf C71} (1996) 179.
\bibitem{alephtwo} F. Ragusa, for the ALEPH coll., talk at the 
			LEPC Meeting, November 19, 1996. 
\bibitem{alephthree} D. Buskulic {\it et al.}, \PLB{356}{95}{409}. 
\bibitem{giudice} M. Carena, G.F. Giudice, S. Lola and C.E.M. Wagner, 
			CERN--TH/96--352, hep--ph/9612334.  
\bibitem{rparityv} There are numerous papers on this subject.
	A partial list includes: \\
	L.J. Hall and M. Suzuki, \NPB{231}{84}{419};\\
	I.H. Lee, \NPB{246}{84}120;\\ 
	S. Dawson, \NPB{261}{85}{297} 297;\\
	R. Barbieri and A. Masiero, \NPB{267}{86}{679};\\ 
	V. Barger, G.F. Giudice, and T. Han, \PRD{40}{89}{2987};\\
	S. Dimopoulos {\it et al.}, \PRD{40}{89}{2987};\\
	H. Dreiner and G.G. Ross, \NPB{365}{91}{597}.
\bibitem{fff} H. Kawai, D.C. Lewellen, and S.-H.H. Tye,
				\NPB{288}{87}{1};\\
		I. Antoniadis, C. Bachas, and C. Kounnas,
				\NPB{289}{87}{87}.
\bibitem{revamp} I. Antoniadis, J.Ellis, J. Hagelin and D.V. Nanopoulos,
			\PLB{231}{89}{65};\\
	J.L. Lopez, D.V. Nanopoulos and K. Yuan, \NPB{399}{93}{654}.
\bibitem{fny} A.E. Faraggi, D.V. Nanopoulos, and K. Yuan,
				\NPB{335}{90}{347}.
\bibitem{alr} I. Antoniadis, G.K. Leontaris and J. Rizos,
			\PLB{245}{90}{161};\\
		G.K. Leontaris, \PLB{372}{96}{212}.
\bibitem{slm}                   A.E. Faraggi,
				\NPB{387}{92}{239}.
\bibitem{eu} A.E. Faraggi, \PLB{278}{92}{131}.
\bibitem{gcu} A.E. Faraggi, \PLB{302}{93}{202}.
\bibitem{custodial}     A.E. Faraggi, \PLB{339}{94}{223}.
\bibitem{ps}     A.E. Faraggi, \NPB{428}{94}{111}.
\bibitem{pati} J.C. Pati, \PLB{388}{96}{532}. 
\bibitem{df} K.R. Dienes and A.E. Faraggi, \NPB{457}{95}{409}. 
\bibitem{dsw} M. Dine, N. Seiberg and E. Witten, \NPB{289}{87}{585}.
\bibitem{elnone} J. Ellis, J.L. Lopez and D.V. Nanopoulos, 
			\PLB{252}{90}{53};\\
		 G. Leontaris and T. Tamvakis, \PLB{260}{91}{333}.
\bibitem{top} A.E. Faraggi, \PLB{274}{92}{47}; hep--ph/9601332, 
		Nuclear Physics {\bf B}, in press.
\bibitem{lykken} S. Chaudhoury, G. Hockney and J. Lykken, 
			\NPB{461}{96}{357}. 
\bibitem{ibanez} L.E. Iba{\~n}ez {\it et al,\/}\/ \PLB{191}{87}{282};\\
		 D. Bailin, A. Love and S. Thomas, \PLB{194}{87}{385};\\
		 A. Font {\it et al,\/}\/ \NPB{331}{90}{421}.
\end{thebibliography}

\vfill
\eject


\textwidth=7.5in
\oddsidemargin=-18mm
\topmargin=-5mm
\renewcommand{\baselinestretch}{1.3}
\pagestyle{empty}
\begin{table}
\begin{eqnarray*}
\begin{tabular}{|c|c|c|rrrrrrr|c|rr|}
\hline
  $F$ & SEC & $SU(3)_C\times SU(2)_L\times SU(2)_c $&$Q_{C'}$ & $Q_L$ & $Q_1$ & 
   $Q_2$ & $Q_3$ & $Q_{4'}$ & $Q_{5'}$ & $SU(5)_H\times SU(3)_H$ &
   $Q_{6'}$ & $Q_{8''}$ \\
\hline
   $L_1$ & $b_1 \oplus$ & $(1,2,2)$&$-{1\over 2}$ & $0$ & ${1\over 2}$ &
   $0$ & $0$ & $-{1\over 2}$ & $-{1\over 2}$ & $(1,1)$ & $-{2\over 3}$ &
   $0$ \\
   $Q_1$ & $1+\alpha+2\gamma$&$(3,2,1)$&${1\over 6}$&$0$&${1\over 2}$ & 
   $0$ & $0$ & $-{1\over 2}$ & $-{1\over 2}$ & $(1,1)$ & ${4\over 3}$ &
   $0$ \\
   $d_1$ &  & $(\overline 3,1,1)$&$-{1\over 6}$ & $-1$ & ${1\over 2}$ & 
   $0$ & $0$ & ${1\over 2}$ & ${1\over 2}$ & $(1,1)$ & $-{4\over 3}$ &
   $0$ \\
   $N_1$ &  & $(1,1,2)$&${1\over 2}$ & $-1$ & ${1\over 2}$ &
   $0$ & $0$ & ${1\over 2}$ & ${1\over 2}$ & $(1,1)$ & ${2\over 3}$ &
   $0$ \\
   $e_1$ &  & $(1,1,2)$&${1\over 2}$ & $1$ & ${1\over 2}$ & 
   $0$ & $0$ & ${1\over 2}$ & ${1\over 2}$ & $(1,1)$ & ${2\over 3}$ &
   $0$ \\
   $u_1$ &  & $(\overline 3,1,1)$&$-{1\over 6}$ & $-1$ & ${1\over 2}$ & 
   $0$ & $0$ & ${1\over 2}$ & ${1\over 2}$ & $(1,1)$ & $-{4\over 3}$ &
   $0$ \\
\hline
   $L_2$ & $b_2 \oplus$ & $(1,2,2)$&$-{1\over 2}$ & $0$&0 & ${1\over 2}$ &
    $0$ & ${1\over 2}$ & $-{1\over 2}$ & $(1,1)$ & $-{2\over 3}$ &
   $0$ \\
   $Q_2$ & $1+\alpha+2\gamma$&$(3,2,1)$&${1\over 6}$&$0$&0&${1\over 2}$ & 
    $0$ & ${1\over 2}$ & $-{1\over 2}$ & $(1,1)$ & ${4\over 3}$ &
   $0$ \\
   $d_2$ &  & $(\overline 3,1,1)$&$-{1\over 6}$ & $1$&0 & ${1\over 2}$ & 
    $0$ & $-{1\over 2}$ & ${1\over 2}$ & $(1,1)$ & $-{4\over 3}$ &
   $0$ \\
   $N_2$ &  & $(1,1,2)$&${1\over 2}$ & $-1$ & 0 & ${1\over 2}$ &
    $0$ & $-{1\over 2}$ & ${1\over 2}$ & $(1,1)$ & ${2\over 3}$ &
   $0$ \\
   $e_2$ &  & $(1,1,2)$&${1\over 2}$ & $1$&0 & ${1\over 2}$ & 
    $0$ & $-{1\over 2}$ & ${1\over 2}$ & $(1,1)$ & ${2\over 3}$ &
   $0$ \\
   $u_2$ &  & $(\overline 3,1,1)$&$-{1\over 6}$ & $-1$&0 & ${1\over 2}$ & 
    $0$ & $-{1\over 2}$ & ${1\over 2}$ & $(1,1)$ & $-{4\over 3}$ &
   $0$ \\
\hline
   $L_3$ & $b_3 \oplus$ & $(1,2,2)$&$-{1\over 2}$ & $0$&0&0 & ${1\over 2}$ &
   $0$ & ${1}$ & $(1,1)$ & $-{2\over 3}$ &
   $0$ \\
   $Q_3$ & $1+\alpha+2\gamma$&$(3,2,1)$&${1\over 6}$&$0$&0&0&${1\over 2}$ &
    $0$ & ${1}$ & $(1,1)$ & ${4\over 3}$ &
   $0$ \\
   $d_3$ &  & $(\overline 3,1,1)$&$-{1\over 6}$ & $1$&0&0 & ${1\over 2}$ &
    $0$ & $-{1}$ & $(1,1)$ & $-{4\over 3}$ &
   $0$ \\
   $N_3$ &  & $(1,1,2)$&${1\over 2}$ & $-1$&0&0 & ${1\over 2}$ &
    $0$ & $-{1}$ & $(1,1)$ & ${2\over 3}$ &
   $0$ \\
   $e_3$ &  & $(1,1,2)$&${1\over 2}$ & $1$&0&0 & ${1\over 2}$ &
    $0$ & $-{1}$ & $(1,1)$ & ${2\over 3}$ &
   $0$ \\
   $u_3$ &  & $(\overline 3,1,1)$&$-{1\over 6}$ & $-1$&0&0 & ${1\over 2}$ &
    $0$ & $-{1}$ & $(1,1)$ & $-{4\over 3}$ & 
   $0$ \\
\hline
\end{tabular}
\label{matter1}
\end{eqnarray*}
\caption{Three generations of massless states and their quantum numbers in the
model of Ref. [13].} 
\end{table}

\end{document}